\documentclass[parskip=half]{scrartcl}

\usepackage[utf8]{inputenc}
\usepackage[english]{babel}
\usepackage{graphicx}
\usepackage{enumitem}
\usepackage{amsmath}
\usepackage{amssymb}
\usepackage{wasysym}
\usepackage{amsthm}
\usepackage[pdftex]{hyperref}

\usepackage{cite}
\usepackage{mciteplus}
\usepackage{xspace}
\usepackage{color}
\usepackage[nottoc]{tocbibind}
\usepackage{pdfpages}

\newcommand{\lsim}{\,{\buildrel < \over {_\sim}}\,}

\newcommand{\pp}         {pp\xspace}

\newcommand{\pbpb}       {\mbox{Pb--Pb}\xspace}
\newcommand{\rts}        {\ensuremath{\sqrt{s}}\xspace}
\newcommand{\rtsnn}      {\ensuremath{\sqrt{s_\mathrm{NN}}}\xspace}
\newcommand{\gevc}       {\ensuremath{\mathrm{\, GeV}\!/c}\xspace}
\newcommand{\gevcc}       {\ensuremath{\mathrm{\, GeV}\!/c ^2}\xspace}
\newcommand{\mev}       {\ensuremath{\mathrm{\, MeV}}\xspace}

\newcommand{\gev}        {\ensuremath{\mathrm{\, GeV}}\xspace}
\newcommand{\tev}        {\ensuremath{\mathrm{\, TeV}}\xspace}
\newcommand{\pt}         {\ensuremath{p_\mathrm{T}}\xspace}

\newcommand{\RAA}  {\ensuremath{R_{AA}}\xspace}
\newcommand{\jpsi}{{\rm J/}\psi}
\newcommand{\dNdphi}{{\rm d}N/{\rm d}\varphi}
\newcommand{\DDbar}{${\rm D}\overline{\rm D}$}
\newcommand{\BBbar}{\ensuremath{{\rm B}\overline{\rm B}\xspace}}
\def\corr{\langle \Delta p_{t,1}, \Delta p_{t,2}\rangle}

\newcommand{\DtoKpi}{\ensuremath{{\rm D}^0 \to {\rm K}^-\pi^+}\xspace}
\newcommand{\DbartoKpi}{\ensuremath{\overline{{\rm D}^0} \to {\rm K}^-\pi^+}\xspace}
\newcommand{\DtoKpipi}{\ensuremath{{\rm D}^+\to {\rm K}^-\pi^+\pi^+}\xspace}

\newcommand{\Dzero}{{\ensuremath{\rm D^0}}\xspace}

\newcommand{\Dstar}{\ensuremath{{\rm D^{*+}}}\xspace}

\newcommand{\Dplus}{\ensuremath{{\rm D^+}}\xspace}

\newcommand{\mur} {\ensuremath{\mu_{\rm R}}\xspace}
\newcommand{\muf}{\ensuremath{\mu_{\rm F}}\xspace}
\newcommand{\muz}{\ensuremath{\mu_{\rm 0}}\xspace}
\newcommand{\mc}{\ensuremath{m_{\rm c}}\xspace}
\newcommand{\lqcd}{\ensuremath{\Lambda_{\rm QCD}}\xspace}

\newcommand{\sigmatot}{\ensuremath{\sigma_{c \overline{c}}}\xspace}
\newcommand{\sigmad}{\ensuremath{\sigma_{D}}\xspace}

\newcommand \Pv{\ensuremath{P_\mathrm{v}}\xspace}

\numberwithin{figure}{section}
\numberwithin{table}{section}
\numberwithin{equation}{section}

\begin{document}
\begin{titlepage}
\begin{center}
\thispagestyle{empty}

\Large\textbf{Habilitationsschrift}\\
\normalsize
\vspace{0.5cm}
zur Erlangung der Venia Legendi\\
f\"ur das  Fach Physik\\
der Ruprecht-Karls-Universit\"at\\
Heidelberg\\
\vspace{5.0cm}
vorgelegt von\\
\Large\textbf{Kai Oliver Schweda}\\
\normalsize
aus Heidelberg\\
\vspace{1.0cm}
\Large\textbf{2013}
\normalsize
\newpage\newpage

\thispagestyle{empty}
\Large\textbf{Prompt production of D mesons with ALICE at the LHC}
\end{center}
\normalsize
\end{titlepage}
\newpage\newpage
\pagenumbering{roman}
\tableofcontents
\newpage
\pagenumbering{arabic}
\section{Introduction}
\label{section:introduction}
The Quark-Gluon Plasma (QGP) is the state of deconfined and thermalized QCD matter produced in high-energy nucleus-nucleus collisions. Its detailed characterization is a major long-term goal of the three large scale experiments ATLAS, CMS and ALICE using proton-proton, proton-lead and lead-lead collisions at unprecedented high energies provided by the Large Hadron Collider.
In the standard Big Bang model, this QGP is the state of matter that permeated the early universe after the electro-weak phase transition, i.e. from picoseconds to about 10 microseconds after the Big Bang. A precise determination of its properties including critical temperature, degrees of freedom, speed of sound, and, in general, transport coefficients would be a major achievement, bringing a far better understanding of QCD as a genuine multi-particle theory. 

\begin{figure}[hbtp]
  \centering
  \includegraphics[width=0.53\textwidth]{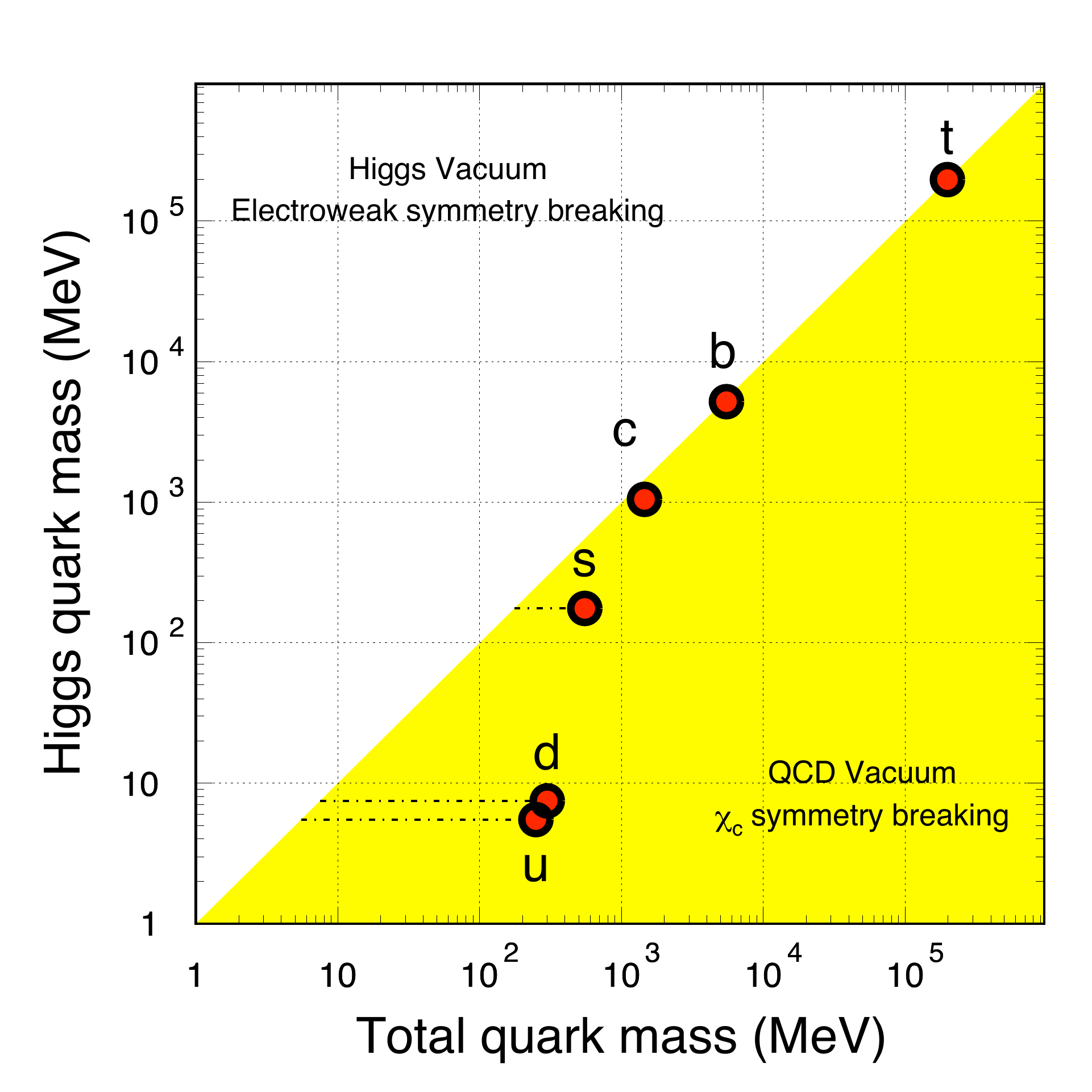}
  \caption{Quark masses in the QCD vacuum and the Higgs vacuum. A large fraction
of the light quark masses arises due to chiral symmetry breaking in the
QCD vacuum. This figure has been taken from~\cite{Zhu}.}
  \label{fig:mass}
\end{figure} 
ALICE successfully started data taking with Pb-Pb interactions in 2010. The experimental data set presently available, already allows for quantitative 
studies of rare probes, i.e. heavy-flavor particles, quarkonia, real and virtual photons, jets, and their correlations with other probes. 

Heavy-flavor particles containing a heavy quark (charm or beauty) are unique probes for studies into QGP bulk properties. Their heavy mass constitutes a new scale in the system that
is much larger than the QCD scale, $\mc, m_{\rm b} \gg \lqcd$, making their production cross sections accessible to calculations in perturbative QCD. 
Their mass is also much larger than the maximum initial QGP temperature.  Heavy quarks acquire their mass almost entirely from
the electroweak sector due to their
coupling to the Higgs field, see Fig.~\ref{fig:mass}. Therefore, they remain heavy even when chiral symmetry is at least partially restored in a QGP. This makes heavy quarks a calibrated probe
for studies of QGP bulk properties. They provide access to the degree of thermalization among quarks and gluons in the QGP. Heavy-quark elliptic flow is especially sensitive to the partonic equation of state, relating macroscopic QGP properties such as energy density, temperature and pressure. Ultimately, heavy quarks might fully equilibrate and become part of the strongly-coupled medium. Studying the in-medium heavy-quark energy-loss mechanism provides both a test and proving ground for the multi-particle aspects of QCD and a probe of the QGP density. In particular, it is crucial to characterize the dependences of energy loss on the parton color charge, mass, and energy, as well as on the density of the medium. 

%
\begin{figure}[hbtp]
  \centering
  \includegraphics[width=0.55\textwidth]{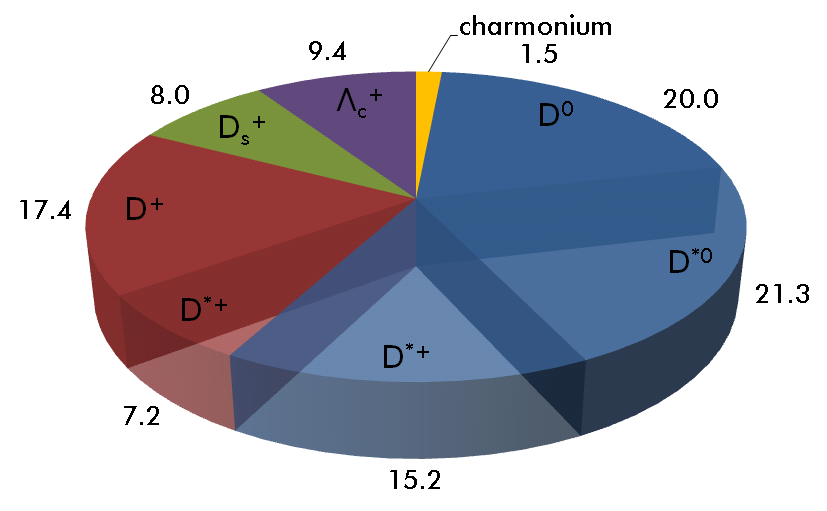}
  \caption{Relative abundances of  charm quarks hadronizing into particular hadron species in vacuo. The contribution from the first excited states $\rm{D}^*(2007)^0$ and
   $\rm{D}^*(2010)^+$ to neutral (blue) and charged (red) D mesons is also indicated~\cite{pdg,jorge}. The numbers given are in units of percent.}
  \label{fig:cake}
\end{figure} 
Experimental studies require the detection of charmed hadrons. The relative abundances of charm quarks hadronizing into a particular hadron species in vacuo are given in
Fig.~\ref{fig:cake}. The contributions from the first excited states $\rm{D}^*(2007)^0$ and $\rm{D}^*(2010)^+$ to neutral and charged D mesons are also indicated. The contribution from higher resonances is less than 5\%, and is neglected in the following. Throughout this thesis, `prompt' refers to all D mesons not originating from weak decays such as $B \rightarrow D+ X$. In particular, ground-state D mesons originating from the decay of excited charm resonances 
also belong to prompt production.

\begin{figure}[hbtp]
  \centering
  \includegraphics[width=0.3\textwidth]{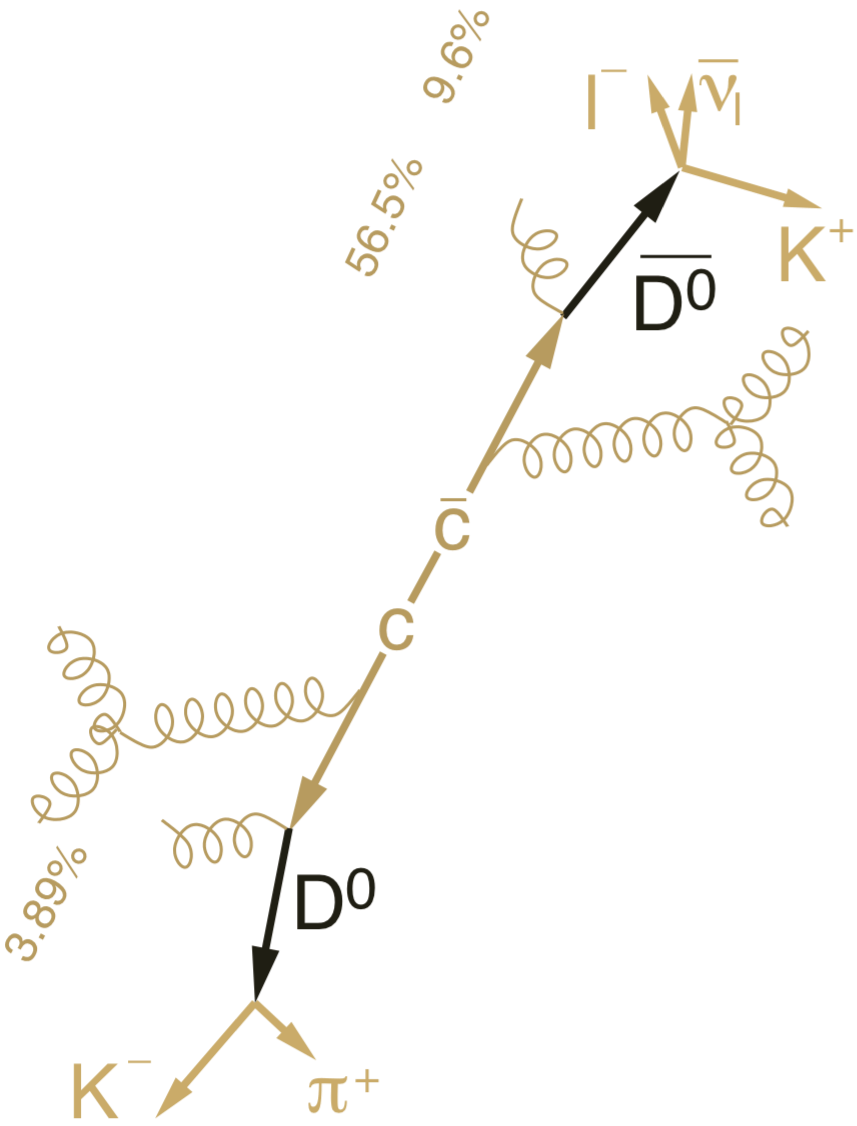}
  \caption{Schematic view of charm-quark hadronization and decay~\cite{pdg,tlusty}. The numbers indicate the probability of a charm quark hadronizing
  into a neutral D meson (56.5\%) and the branching ratios for the pictured hadronic (3.89\%) and semi-leptonic (9.6\%) decay channels.}
  \label{fig:decay}
\end{figure} 
Charmed hadrons decay before they reach any active detector material. Thus they have to be reconstructed from their decay products. The production and decay of a charmed hadron is schematically shown in Fig.~\ref{fig:decay}. A charm and an anti-charm quark are created 
in pairs
in strong interactions, mostly due to gluon-gluon processes. Both charm quarks hadronize  e.g. into a neutral D and $\overline{\rm {D}}$ meson, which decay at the
secondary vertex after moving typically a fraction of a millimeter away from the primary collision vertex. 
\begin{table}
\centering
\begin{tabular}{lcclc} 
\hline \hline
meson & mass\, ($\mathrm{MeV}\!/c ^2$) & $c\tau$\, ($\mu$m) or $\Gamma$\, (keV) & decay mode & branching ratio\, (\%) \\
\hline
\Dzero & $1864.86 \pm 0.13$ & $122.9  \pm 0.4$ & ${\rm K}^-\pi^+$ &  $3.88\pm 0.05$\\
&&& ${\rm K}^+\pi^-$ & $(1.47 \pm 0.07) \times 10^{-2}$ \\
\Dplus & $1869.62 \pm 0.15$ & $311.8 \pm 0.2$ &  ${\rm K}^-\pi^+\pi^+$ & $9.13\pm 0.19$\\
${\rm D}_s^+$ & $1968.50 \pm 0.32$ & $149.9 \pm 2.1$ & ${\rm K}^- \rm{K}^+ \pi^+$ & $5.49 \pm 0.27$\\
$\rm{D}^*(2007)^0$ & $2006.99 \pm 0.15$ & $<2100$ & $\Dzero \pi^0$ & $61.9\pm 2.9$\\
&&&$\Dzero + \gamma$ & $38.1 \pm 2.9$\\
$\rm{D}^*(2010)^+$ & $2010.29 \pm 0.13$ & $83.3 \pm 1.2 \pm 1.4$ & $\Dzero \pi^+$ & $67.7\pm 0.5$\\
&&&$\Dplus \pi^0$ & $30.7 \pm 0.5$\\
\hline \hline
\end{tabular}
\caption{Some properties of D mesons~\cite{pdg,Lees:2013uxa}. For the charmed resonances D$^*$, the natural line width $\Gamma$ is given. }
\label{tab:decay}
\end{table}

The detection of charmed hadrons through their semi-leptonic decays has the advantage of larger branching ratios and the possibility of triggering on high-momentum leptons. However, due to the decay kinematics, information on the charmed hadron momentum is largely lost~\cite{Batsouli}. Reconstructing charmed hadrons in their hadronic decay channels retains the full kinematic information,
thus giving direct access to the modification of charmed hadron production in the QGP medium. Some properties of charmed hadrons and their decay modes are listed in Tab.~\ref{tab:decay}.

The hadrons's invariant mass is calculated using information from the detected decay daughter candidates.
The charmed hadron yield then appears as a peak at the rest mass in the invariant mass distribution above a combinatorial background of uncorrelated particles. In order to enhance the signal-to-background ratio, excellent particle identification capabilities at low to intermediate
momentum and precise sub-millimeter pointing resolution to the primary vertex are essential. In the ALICE apparatus~\cite{Aamodt:2008zz}, this is provided by the Time Projection Chamber (TPC)~\cite{TPC}, which performs 3-dimensional tracking of charged particles and provides information on the specific energy deposit ${\rm d}E/{\rm d}x$ in the TPC gas mixture. The resolution of the ${\rm d}E/{\rm d}x$ measurement is close to the theoretical limit, i.e. it is dominated by intrinsic fluctuations of the energy deposit of charged particles
when propagating in the TPC gas~\cite{TPC}. Particle identification is extended to intermediate momentum by the Time-of-Flight detector (TOF)~\cite{TOF}, which uses multi-gap resistive plate chambers with an overall time resolution of $90\, {\rm ps}$ to separate kaons from
pions on a track-by-track basis up to a momentum of $1.5\gevc$.  Ultra-high spatial resolution is provided by the Inner Tracking System (ITS),
which is based 
on silicon detector technology. Two inner layers of high-granularity active pixel sensors measure the daughter particles
at a radial distance of  $3.9\, {\rm cm}$
 away from the primary vertex with an intrinsic resolution of $12\, \mu {\rm m}$ in the transverse plane, resulting in a pointing resolution of $75\, \mu {\rm m}$ at the primary vertex for momenta larger than $1\gevc$~\cite{ITSalign}. 

\begin{figure}[hbtp]
  \centering
  \includegraphics[width=0.99\textwidth]{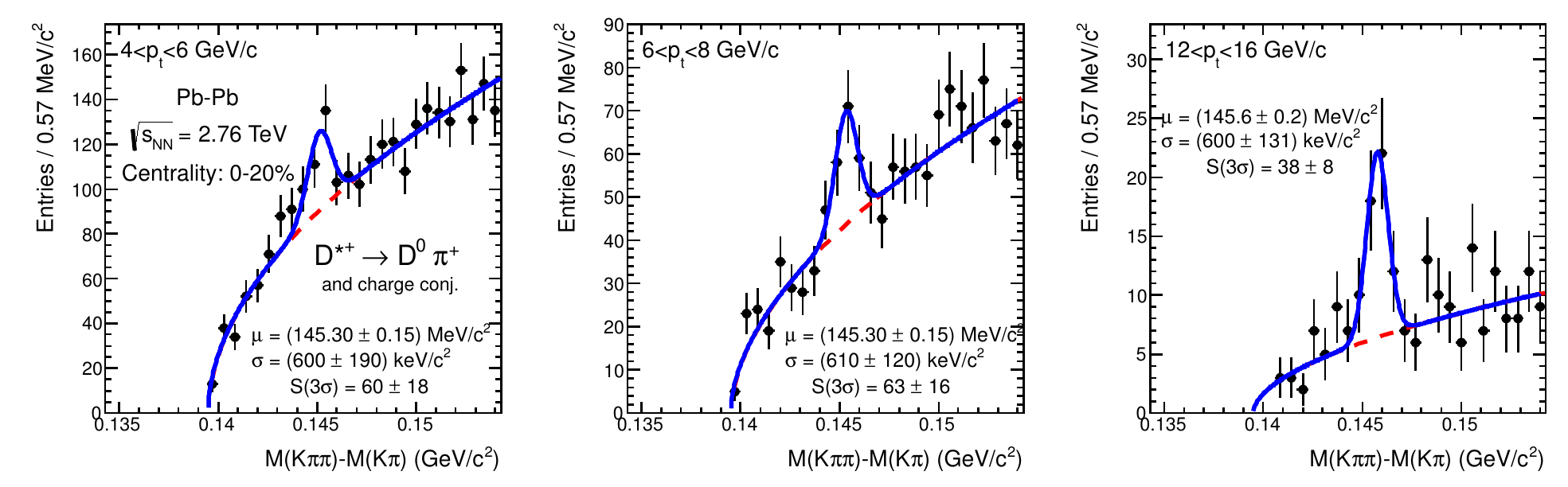}
  \caption{Invariant mass distribution of $\rm{D}^*(2010)^+$ candidates from 20\% most central \pbpb collisions
  at \rtsnn = 2.76~TeV for different momentum bins. This figure has been taken from~\cite{dv2}.}
  \label{fig:inv-mass}
\end{figure} 

The invariant mass distribution of $\rm{D}^*(2010)^+$ candidates from 20\% most central \pbpb collisions
at \rtsnn = 2.76~TeV for different momentum bins is shown in Fig.~\ref{fig:inv-mass}.
Note that the mass difference between the invariant mass of reconstructed $\rm{D}^*(2010)^+$ and $\rm{D}^0$ candidates is shown.  
Due to the
small phase space available for the kinetic energy of the soft pion originating from the $\rm{D}^*(2010)^+$ decay, the signal appears
as a narrow peak at low invariant mass, close to the kinematic threshold. This leads to a small contribution from the combinatorial background.
Since the daughter D meson is much heavier than the soft pion, in this representation the mass resolution is solely given by the momentum resolution of the soft pion.
This results in a narrow width of the signal peak of less than $1\, {\rm MeV}$, compared to typically $20\, \rm{MeV}$ for the other D mesons. Both the rather small 
contribution from the combinatorial background and the narrow signal peak make the statistical significance of the \Dstar measurement practically 
identical to the $D^0$ measurement, with a higher momentum reach for the \Dstar. 

In this thesis, results from three decay modes for D mesons are reported, namely $\DtoKpi$, with a branching ratio
of $(3.88\pm 0.05)\%$; $\DtoKpipi$, with a branching ratio of $(9.13\pm 0.19)\%$; 
\mbox{$\rm{D}^*(2010)^+\to \Dzero \pi^+$}, which strongly decays to \Dzero with a branching ratio of $(67.7\pm 0.5)\%$, and subsequently 
follows the channel $\DtoKpi$;
and  their charge conjugates~\cite{pdg}. Strictly speaking, the invariant mass distribution for the decay $\DtoKpi$ consists of the Cabibbo-favored
decay \DtoKpi and the doubly Cabibbo-suppressed decay \DbartoKpi. The latter has a branching ratio of $(1.47 \pm 0.07) \times 10^{-4}$, and has been neglected. 
All cross sections reported are the sum for each D meson and its anti-particle divided by two. Contributions from the weak decays of B mesons are momentum-dependent, and amount to roughly 10 - 15\%. These contributions have been subtracted by using results from state-of-the-art calculations in perturbative QCD. Results for $\rm{D}_s^+$ production in proton-proton collisions can be found in \cite{Ds}. 
Recent reviews of heavy-flavor production at collider energies and their implications for the properties of QCD matter are given in~\cite{Averbeck:2013oga,Rapp:2008tf}. 

This thesis is organized as follows. Results on charmed hadron production in proton-proton collisions with ALICE are presented in Sect.~\ref{section:pp}, and results from lead-lead collision are presented in Sect.~\ref{section:pbpb}. Model studies on charm anti-charm correlations and their implications on thermalization in high-energy nucleus-nucleus collisions are discussed in Sect.~\ref{section:model}. An outlook is given in Sect.~\ref{section:outlook}.
\newpage
\section{Results}
\label{section:results}

\subsection{Prompt production of D mesons in \pp collisions}
\label{section:pp}

The measurement of charm and beauty production cross sections in proton--proton (\pp) collisions at the Large Hadron Collider (LHC) constitutes a challenge to our understanding of calculations in perturbative Quantum Chromo--Dynamic (pQCD) at the highest  collider energies.  
These calculations use the factorization approach to describe heavy-flavor hadron production as a convolution of three terms: the parton distribution function, the hard parton scattering cross section and the fragmentation function. The parton distribution function describes the initial distribution 
in Bjorken-x of quarks and gluons in the colliding protons. The hard parton scattering cross section is calculated as a perturbative series in 
terms of 
the coupling constant of the strong interaction. The fragmentation function parametrizes the relative production yield and momentum distribution for a heavy quark that hadronizes into a particular hadron species.

The production cross section of beauty hadrons at Tevatron at a collision energy of $\rts=1.96\tev$~\cite{cdfB,fonllBcdf,gmvfnsBcdf} and at the LHC at $\rts=7\tev$~\cite{lhcbBeauty,cmsJpsi} is well described by perturbative calculations at next-to-leading order, e.g. GM-VFNS~\cite{gmvfns1,gmvfns2}, or at fixed order with a next-to-leading-log resummation, i.e. FONLL~\cite{fonll1, fonll2}. 
Charmed hadron production at Tevatron~\cite{charmcdf,fonllDcdf,gmvfnsDcdf}, at the LHC~\cite{aliceDmeson7TeV,ATLASDmeson,LHCbDmeson}, and  RHIC, e.g. the heavy-flavor decay lepton measurements at $\rts=200\gev$~\cite{phenixelepp,starelepp}, are also reproduced by the pQCD calculations within experimental and theoretical uncertainties. 
\begin{figure}[htb]
  \centering
  \includegraphics[width=0.5\textwidth]{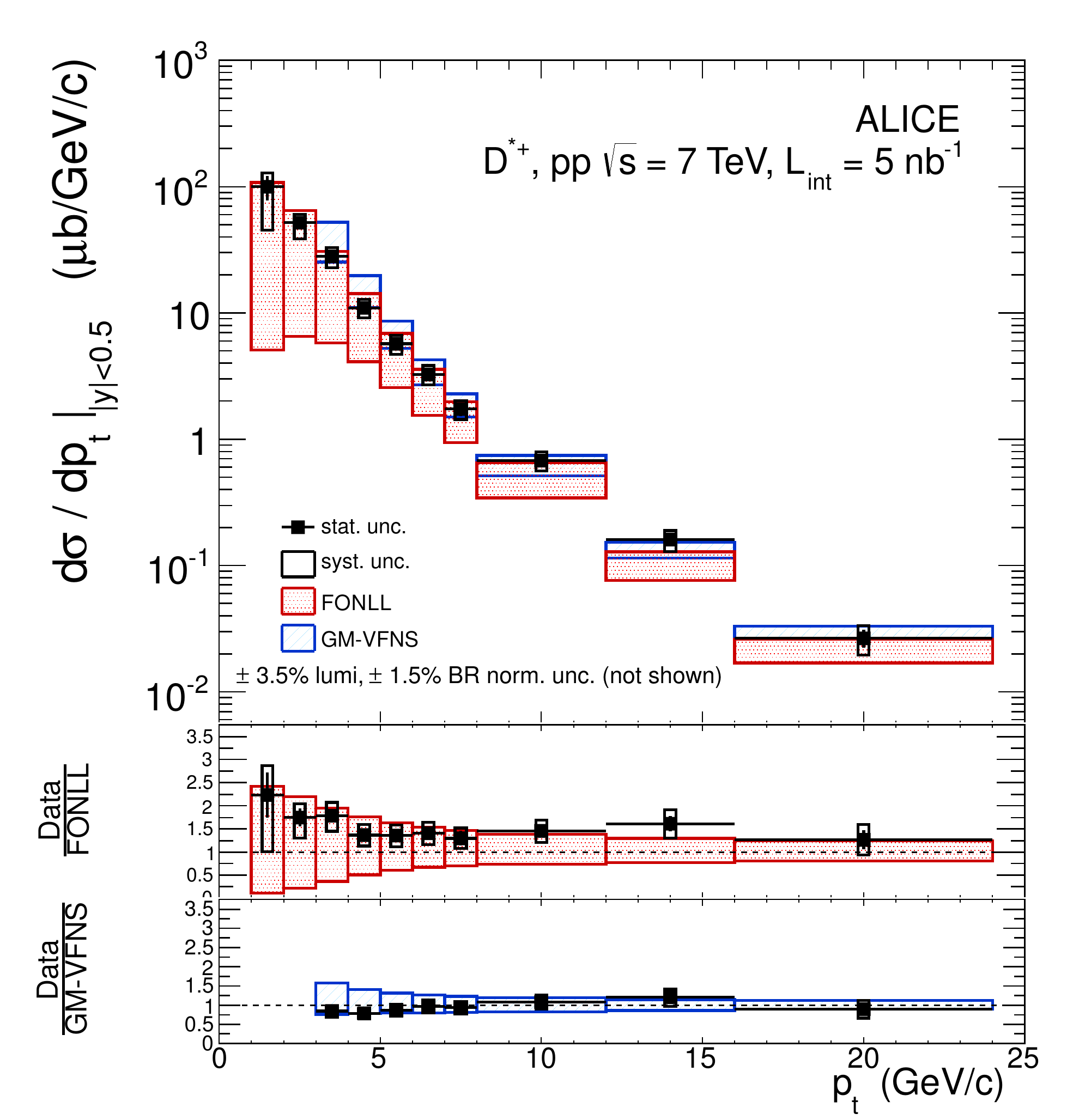}
  \caption{Cross section for prompt production of $\Dstar$ mesons in pp collisions at $\rts=7\tev$ and predictions from calculations in perturbative QCD within  the FONLL~\cite{FONLLalice}~(red) 
and the GM-VFNS~\cite{VFNSalice}~(blue) framework. This figure has been taken from~\cite{aliceDmeson7TeV}.}
  \label{fig:pp-spectrum}
\end{figure} 
%
%
On the other hand, the relative abundances of charmed hadrons test the statistical hadronization model~\cite{shm-charm} of charm quarks forming hadrons.
If the hadronization of a charm quark occurs statistically, it should be independent of the collision system and energy. 
%
Finally, heavy-quark production rates in \pp~collisions provide an essential baseline for studies into the bulk properties of the 
Quark Gluon Plasma~\cite{stathadronization,hqcolorimetry}.


The cross section for prompt production of $\Dstar$ mesons in pp collisions at $\rts=7\tev$ is shown in Fig.~\ref{fig:pp-spectrum} as a function of transverse momentum. The spectrum starts at 1\gevc and extends up to 24\gevc. At lower momenta, D mesons do not move far enough away from the primary vertex before the charm quark decays, and thus do not survive the topological selection. At the highest momenta, the invariant mass spectrum is essentially free of background and the measurement is solely restricted by statistics.   
Predictions from calculations in perturbative QCD within the FONLL~\cite{FONLLalice}~(red)
and the GM-VFNS~\cite{VFNSalice}~(blue) framework are shown by shaded areas. Both calculations use the CTEQ6.6~\cite{cteq6.6} 
next-to-leading-order parton distribution functions as input. The systematic uncertainties in each calculation scheme was estimated by 
simultaneously varying the factorization scale variable \muf and the renormalization scale variable \mur around their central value \muf = \mur = \muz = $\sqrt{\pt^2 +\mc^2}$  by a  factor of two with the constraint $0.5 \le \muf / \mur \le 2$. In the FONLL calculations, the charm quark mass was also varied around the central value of
$\mc =1.5\gevcc \pm 0.2\gevcc$. Results from both calculations are able to describe the measurement within rather large uncertainties. Interestingly, all data points populate the upper band of results from FONLL. This behavior has already been observed at lower collision energies and different collision systems. 
 On the other hand, compared to results from GM-VFNS, the data points sit on the lower side, which is at odds with observations at Tevatron energies.
 This indicates a steeper energy dependence in the GM-VFNS framework than in data. 
\begin{figure}[hbt]
  \centering
  \includegraphics[width=0.5\textwidth]{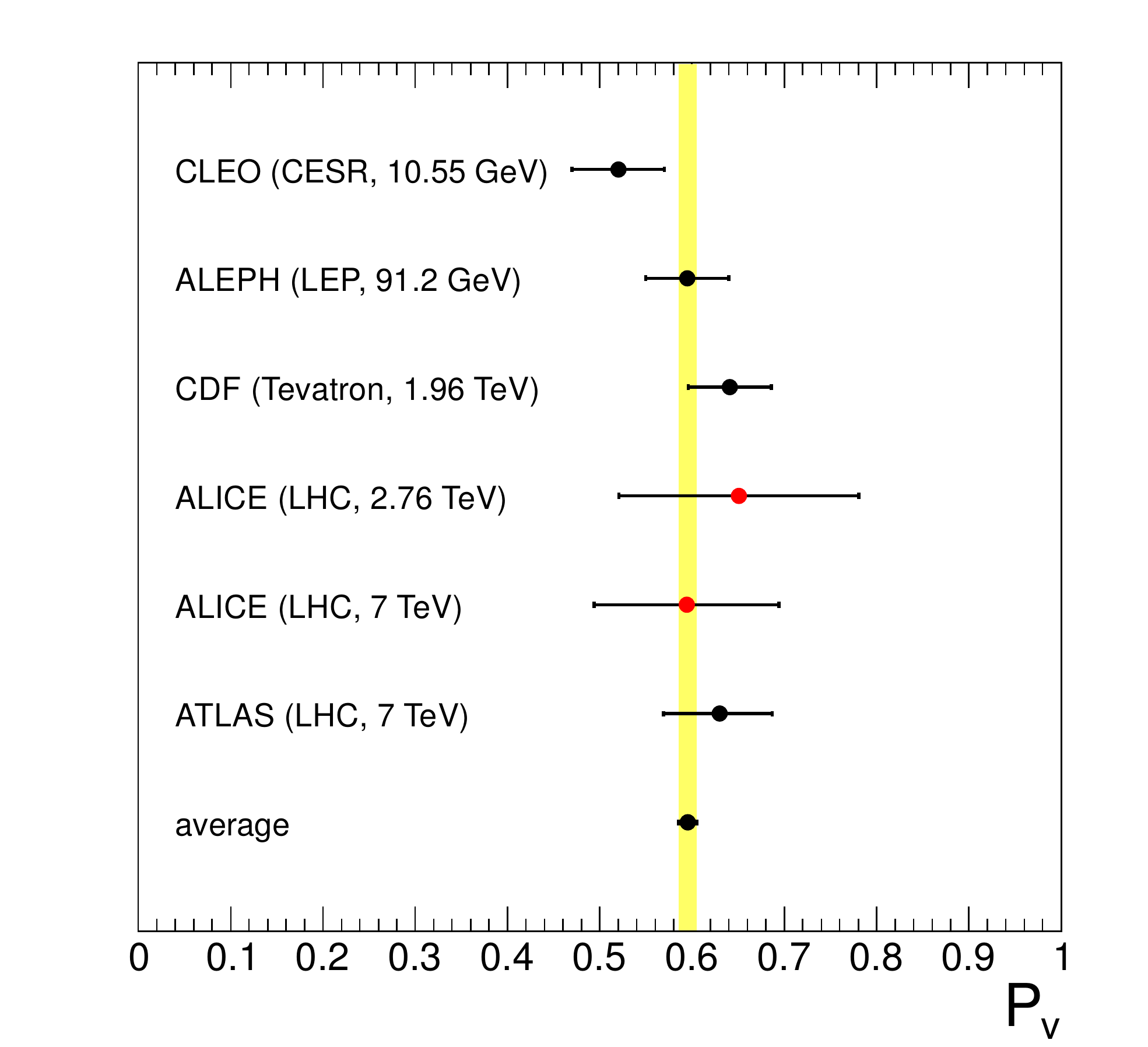}
  \caption{The fraction $\Pv$ of D mesons  with valence quark content $|c \overline{d} \rangle$ created in a vector state over all D mesons with the same
  valence quark content~\cite{CLEO,ALEPH,charmcdf,CDFPv,STARDmeson,aliceDmeson7TeV,ATLASDmeson}. The weighted average of the experimental measurements reported in Ref.~\cite{ADavid} and of the LHC data~\cite{aliceDmeson7TeV,ATLASDmeson} shown in the figure is $\Pv = 0.60 \pm 0.01$, and is represented by a solid yellow vertical band. This figure has been taken from~\cite{Dpp2.76}.}
  \label{fig:pv}
\end{figure} 

In general, experiments only cover a small fraction of the full phase space populated by charm production. For example, with the present limitation in $\pt > 1\gevc$, ALICE covers about 70-80\% of the transverse momentum spectrum at mid-rapidity and roughly 10\% of the total production cross section
for each D meson.
Thus extrapolating the data by simply adding results from calculations in pQCD in the uncovered space phase would lead to results that are 
largely
dominated by the calculations. Therefore, the relative distribution, i.e. the shape in transverse momentum and rapidity, is taken from the pQCD calculations to determine the fraction of charm production covered by the experiment. 
In the following, the default extrapolation method used by all high-energy experiments is described. 
For each D meson species, the total production cross section $\sigma_{\rm D}$ was extracted by multiplying the measured cross section by the ratio of the calculated total cross section over the calculated cross section in the experimentally covered phase space. 
The systematic uncertainties of the calculation were estimated by varying the renormalization ($\mur$) and factorization ($\muf$) scale variables and the charm quark mass ($\mc$), as described above. 
Uncertainties in the parton distribution functions were estimated using the CTEQ6.6~\cite{cteq6.6} PDF uncertainty eigenvectors  and adding the largest positive and negative variation in quadrature. Finally, all three contributions were added in quadrature.

One way of addressing charm quark hadronization is to consider the ratio of D mesons with valence quark content $|c \overline{d} \rangle$ created in a vector state (spin 1) to those produced in a vector (spin 1) or a pseudoscalar state (spin 0). This ratio, $\Pv$, is calculated 
by taking the ratio of $\sigma_{\Dstar}$ over the sum of $\sigma_{\Dstar}$ and the part of  $\sigma_{\Dplus}$ not originating from $\Dstar$ decays,
 \begin{equation}
\Pv
= \frac{\sigma_{\Dstar}}{\sigma_{\Dstar}+\sigma_{\Dplus}
- \sigma_{\Dstar} \cdot (1 - {\rm BR}_{\Dstar\rightarrow {\rm D}^0\pi^+})}
= \frac{\sigma_{\Dstar}}{\sigma_{\Dplus}+\sigma_{\Dstar}\cdot {\rm BR}_{\Dstar\rightarrow {\rm D}^0\pi^+}}.
\label{eq:extrap_pv}
\end{equation}
The advantage of this representation is that the right-hand side of Eq.(\ref{eq:extrap_pv}) only contains one branching ratio with a small experimental uncertainty, while for the decay 
to $\Dplus$ two branching ratios either including a neutral pion or a photon would be required.  

With the ALICE data, this results in 
\begin{eqnarray}\hspace*{-2in}
\Pv (\mathrm{2.76\tev}) = 0.65 \pm 0.10 \, (\mathrm{stat.}) \,  \pm 0.08 \, (\mathrm{syst.} ) \, \pm 0.002 \, (\mathrm{BR}) \, ^{+0.011}_{-0.003} (\mathrm{extr.})\, , \nonumber
\\
\Pv (\mathrm{7\tev}) = 0.59 \pm 0.06 \, (\mathrm{stat.}) \, \pm 0.08 \, (\mathrm{syst.} ) \, \pm 0.002 \, (\mathrm{BR}) \, ^{+0.005}_{-0.002} (\mathrm{extr. }) \, , \nonumber
\end{eqnarray}
where uncertainties due to the extrapolation into the full phase space and branching ratios are negligible.
These values coincide with results from other experiments at different collision energies and for different colliding systems~\cite{CLEO,ALEPH,charmcdf,CDFPv,STARDmeson,aliceDmeson7TeV,ATLASDmeson}, as shown in Fig.~\ref{fig:pv}. 
This means that the charm quark does not remember its origin and hadronizes independently of its production mechanism. 
In hindsight, this justifies the factorization ansatz to describe
charmed-hadron production with the creation of a charm--anti-charm quark pair occurring at an early time scale and its hadronization at a later one. 

The weighted average of the experimental measurements reported in Ref.~\cite{ADavid}, with average $0.594 \pm 0.010$, with the LHC data~\cite{aliceDmeson7TeV,ATLASDmeson} shown in Fig.~\ref{fig:pv}~is $\Pv^{\rm average} = 0.60 \pm 0.01$, which is represented by a solid yellow vertical band in the figure. 
Note that \Pv can also be determined from the ratio of the ground-state D mesons $\Dzero$ and $\Dplus$. This is due to the fact the excited state $D^*(2007)^0$ solely decays to 
\Dzero due to energy conservation and $D^*(2010)^+$ decays to \Dzero and \Dplus with a ratio of roughly 2:1 due to isospin conservation, 
see Fig.~\ref{fig:cake}. This leads to an asymmetry in the prompt production of
\Dzero and \Dplus, i.e. a ratio different from unity.

The expectation from na\"{\i}ve spin counting amounts to $\Pv^{\rm Spin \, counting} = 3/(3+1)=0.75$, which does not agree with data.
The argument of na\"{\i}ve spin counting originates from heavy-quark effective theory (HQET) assuming an infinitely large heavy-quark mass,
meaning that the effect due to the mass difference between $\Dstar$ and $\Dplus$ is negligible.
With a relative mass difference of $\Delta m / m \approx 7.5\%$  in the D meson system, HQET is not a good assumption for the charm quark.
This contrasts with the B meson system which has $\Delta m / m \approx 8.7\permil$.
 
Calculations combining the Lund symmetric fragmentation function with exact Clebsch-Gordan coefficient coupling from the virtual quark--antiquark pair to the final hadron state functions predicts $\Pv^{\rm Lund \, frag} \approx 0.63$~\cite{LundFrag}, which is in good agreement with data. In this model, due to Clebsch-Gordan coefficient coupling, spin counting is automatic, while differences in the hadron mass are taken into account in the fragmentation function by an exponential term. 
On the other hand, in the Statistical Model~\cite{shm-charm,stathadronization}, the ratio of the total yields of the directly-formed charmed mesons $\Dstar$ and $\Dplus$, which have identical valence quark content, is expected to be 
$3 \cdot \left( m_{\Dstar} / m_{\Dplus} \right)^2  \cdot \exp\left(  -  ({m_{\Dstar} - m_{\Dplus} }) / T \right) \approx 1.4$ for a temperature parameter of $T=164\, MeV$, 
where the factor of three comes from spin counting. 
We calculate $\Pv^{\rm Stat. \, Model} \approx 0.58 \pm 0.13 $ for $T=(164\pm10)\, MeV)$. 
Other implementations of the Statistical Model~\cite{becattini,rapp-dmeson} predict similar values of $\Pv$, ranging between 0.55 and 0.64.
The experimental results for \Pv thus allow for a description within the statistical hadronization of charm~\cite{shm-charm,stathadronization} or calculations considering the Lund symmetric fragmentation function~\cite{LundFrag}. 

\begin{figure}[hbt]
  \centering
  \includegraphics[width=0.6\textwidth]{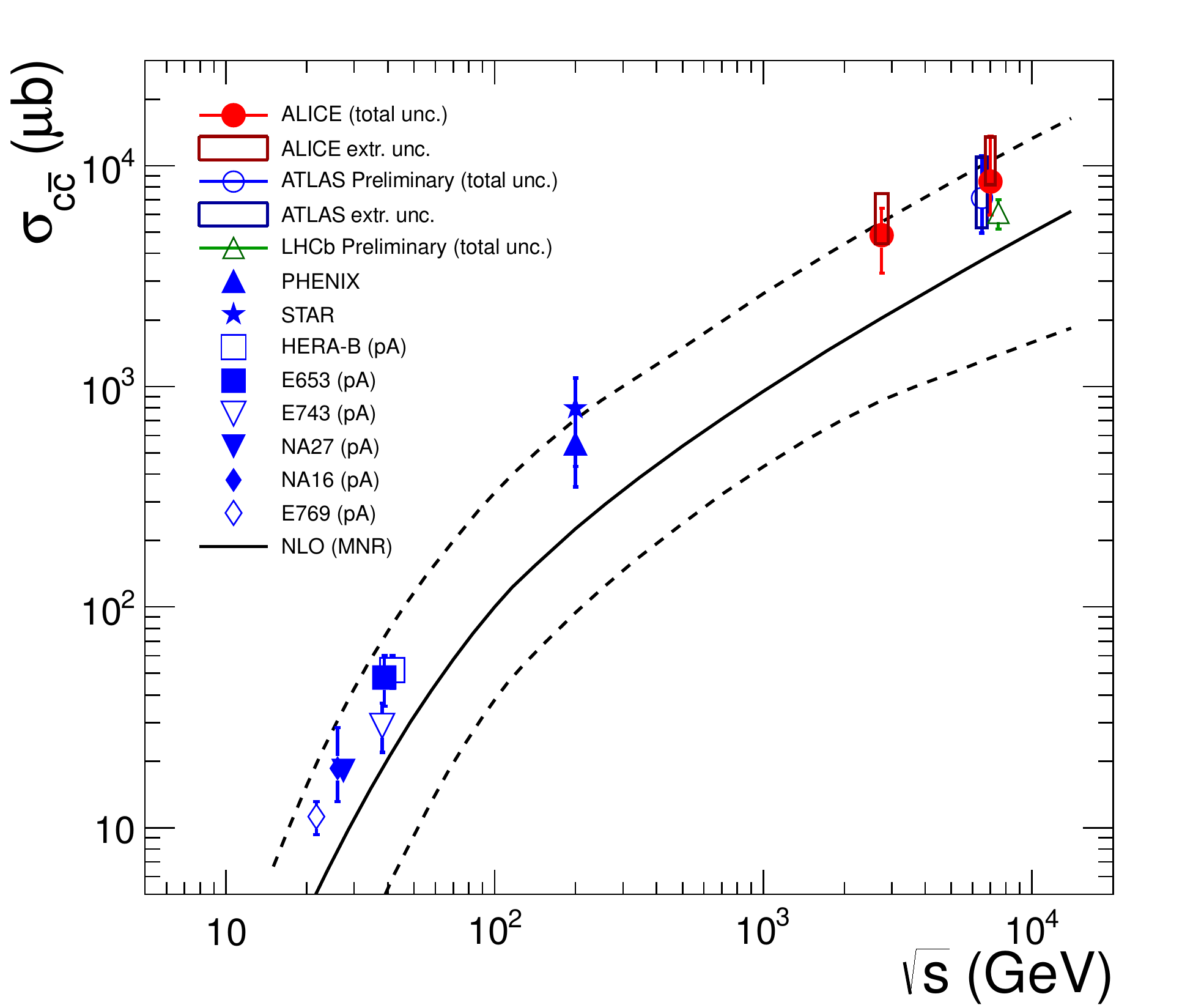}
  \caption{Total charm production cross section  \sigmatot per nucleon-nucleon pair versus collision energy~\cite{Lourenco,STARallCharm,ATLASDmeson,LHCbDmeson,PHENIX}. In case of proton-nucleus (p-A) or deuteron--nucleus (d-A) collisions, the measured cross sections have been scaled down by the number of binary nucleon--nucleon collisions  calculated in a Glauber model of the proton--nucleus or deuteron--nucleus collision geometry. Results from calculation in perturbative QCD~\cite{mnr}~(and their uncertainties) are represented by solid (dashed) lines.
  This figure has been taken from~\cite{Dpp2.76}.}
  \label{fig:cross-section}
\end{figure} 
The total charm production cross section \sigmatot was estimated for each species of D meson separately by dividing the total D meson production cross section \sigmad by the relative production yield for a charm quark hadronizing to a particular species of D meson, that is the fragmentation fractions (FF) of $0.557 \pm 0.023$ for $\Dzero$, $0.226 \pm 0.010$ for $\Dplus$, and $0.238 \pm 0.007$ for $\Dstar$~\cite{pdg}. 
The measured yields are consistent with these ratios.
The weighted average of the total charm production cross section was then calculated from the extrapolated values for  $\Dzero$, $\Dplus$, and  $\Dstar$.

%

The total nucleon-nucleon charm production cross section~\cite{Lourenco,STAR,ATLASDmeson,LHCbDmeson,PHENIX} 
and its dependence of on the collision energy is shown in Fig.~\ref{fig:cross-section}. The uncertainty boxes around the ATLAS~\cite{ATLASDmeson} and ALICE~\cite{aliceDmeson7TeV} points denote the extrapolation uncertainties alone, while the uncertainty bars represent the overall uncertainties. Note that in case of proton--nucleus (p-A) or deuteron--nucleus (d-A) collisions, the measured cross sections have been scaled down by the number of binary nucleon--nucleon collisions  calculated in a Glauber model of the proton--nucleus or the deuteron--nucleus collision geometry. 
At $\rts=7\tev$, results from ALICE and preliminary measurements by the ATLAS~\cite{ATLASDmeson} and the LHCb Collaboration~\cite{LHCbDmeson} are in fair agreement. 
The curves show the calculations at next-to-leading-order within the MNR framework~\cite{mnr} together with their uncertainties using the same parameters (and parameter uncertainties) discussed above for FONLL. 
The dependence on the collision energy is described by pQCD calculations. 
Interestingly, all data points populate the upper band of the theoretical prediction. 
This might hint to a smaller actual charm quark mass than the central value of \mc = 1.5\gevcc assumed in the pQCD calculations. A recent combined next-to-leading order QCD analysis of charm production in
deep inelastic electron-proton scattering at HERA using the $\overline{MS}$ running mass scheme determined a lower charm quark mass  of $\mc \approx 1.26 \gevcc$~\cite{charmHera}.
A smaller charm mass would implicitly lead to a larger cross section of gluons splitting into pairs of charm and anti-charm quarks.  A deficit in gluon splitting processes to the production of charm has been indicated in Monte Carlo calculations when compared to measurements at Tevatron in the associated production of Z bosons with charm quark jets~\cite{Abazov:2013hya} and photon-tagged heavy-quark jets~\cite{D0:2012gw,Aaltonen:2013coa} as well as at the LHC in the measurement of associated charm production in W final states~\cite{CMS:qwa}.

\newpage
\subsection{Prompt production of D mesons in \pbpb collisions}
\label{section:pbpb}

Heavy quarks are almost exclusively produced in the initial stage of a collision in high-virtuality scattering
processes, and their annihilation rate 
is small~\cite{PBM}.
Hence, final state heavy-flavor hadrons at all transverse 
momenta originate from heavy quarks that have experienced all stages of the 
system's evolution.
Due to interactions with the medium, their production yields are sensitive to medium properties such as energy density, temperature, and
in general transport properties.
Heavy quarks lose energy through inelastic processes such as medium-induced gluon radiation~\cite{gyulassy,bdmps}, and elastic processes such as collisions with other partons in the medium~\cite{thoma}.
Since the color-charge factor of quarks is smaller than that of gluons, the
energy loss of quarks should be smaller than for gluons. 
Additionally, the dead-cone effect reduces the available phase space for small-angle gluon radiation 
due to conservation of angular momentum~\cite{Ellis:1991qj}
for heavy quarks at moderate energy-over-mass values~\cite{hqcolorimetry,asw,dg,wang,whdg},
leading to even smaller energy loss. 
However, other proposed mechanisms such as  
in-medium hadron formation and dissociation~\cite{adil,vitev} would lead
to a larger energy loss of heavy-flavor hadrons, characterized by smaller formation 
times than for light-flavor hadrons.
Ultimately, low-momentum heavy quarks might sufficiently interact with the medium constituents to even thermalize and become
part of the medium, e.g. through re-scattering and in-medium resonant 
interactions~\cite{rapp}.

The effects of parton energy loss, or in general medium effects, are quantified in the nuclear modification factor $R_{AA}$:
\begin{equation}
\label{eq:raa}
R_{AA}(p_T) = \frac{1}{\langle N_{\rm coll} \rangle} \cdot \frac{{\rm d}N_{AA} / {\rm d}p_T}{{\rm d}N_{pp} / {\rm d}p_T} =  \frac{1}{\langle T_{AA} \rangle} \cdot \frac{{\rm d}N_{AA} / {\rm d}p_T}{{\rm d} \sigma _{pp} / {\rm d}p_T}. 
\end{equation}
Here, $\langle T_{AA} \rangle$ is the average nuclear overlap function, as calculated in a Glauber model of the nucleus-nucleus collision geometry.
For hard processes, production yields in nucleus-nucleus $(AA)$ collisions are -- in the absence of any nuclear medium effects -- expected to scale with the number $\langle N_{\rm coll} \rangle$ of binary nucleon-nucleon collisions when compared to pp yields. As a consequence, $R_{AA}$ equals unity.

\begin{figure}[hbtp]
  \centering
  \includegraphics[width=0.58\textwidth]{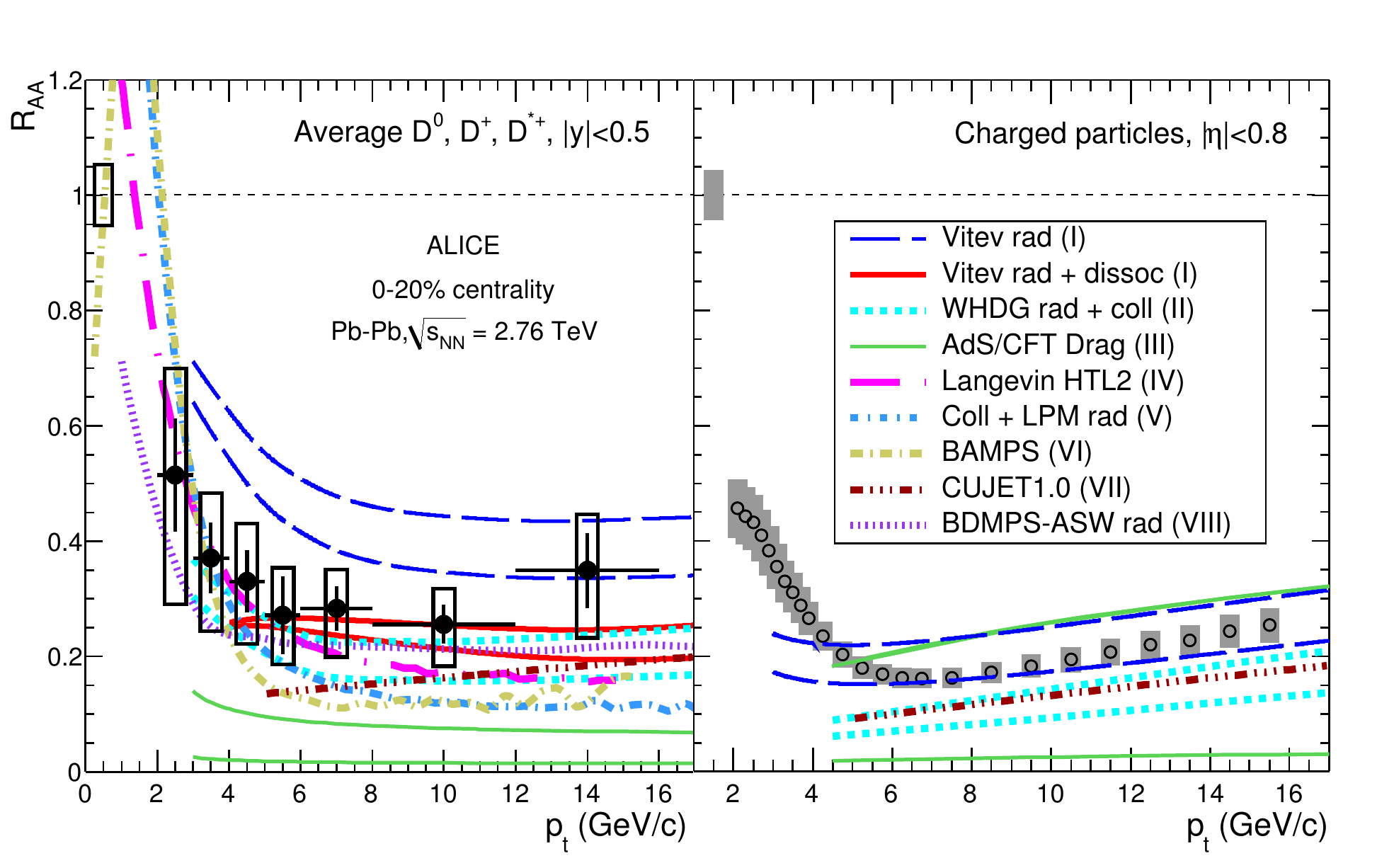}
  \caption{Nuclear modification factor \RAA of D mesons (left) and unidentified charged particles (right) and results from model calculations.
  This figure has been taken from~\cite{aliceDRAA}.}
  \label{fig:raa}
\end{figure} 
Heavy-quark masses are not negligible 
at momenta below $\pt \lsim 10\gevc$.
A mass hierarchy in the nuclear modification factor $\RAA$ value is expected when going from 
mostly gluon-originated light-flavor hadrons, e.g. pions, 
to D and B mesons (see e.g.~\cite{whdg,adsw}), hence 
$\RAA^\pi<\RAA^{\rm D}<\RAA^{\rm B}$. Details depend on the reference spectrum from \pp collisions.
A measurement and comparison of light versus heavy mass probes 
thus provides a unique test of the color-charge and mass dependence of parton energy loss.

The high statistics \pp spectrum was obtained at a higher collision energy of \rts = 7\tev, which was the maximum possible
collision energy in 2010.
The \pp reference spectrum for the $\RAA$ measurements was obtained 
by scaling these results to the lower \pbpb collision energy of \rtsnn=2.76\tev via a pQCD-driven approach. 
These results were validated by comparing to experimental data from a \pp sample of limited statistics taken at the same collision energy~\cite{Dpp2.76}. 
The nuclear modification factor for D mesons, averaged over \Dzero, \Dplus, and \Dstar,  as measured by ALICE, is shown in the right panel of Fig.~\ref{fig:raa}. 
The vertical bars represent statistical uncertainties in central collisions, having a typical magnitude of 20--25\%
for $\Dzero$  and  30--40\% for $\Dplus$ and $\Dstar$ mesons. 
The results for the three D meson species numerically coincide within statistical uncertainties, exhibiting 
a suppression by a factor 3-4, i.e. $\RAA\approx 0.25$--0.3, in central collisions for $\pt>5\gevc$.
For decreasing $\pt$, the nuclear modification factor of $\Dzero$ 
in central collisions tends to exhibit a lower suppression. Since charm is conserved, the depletion of the spectrum at large momentum should correspond to an enhancement at lower momentum. Also, if charm participates in the collective expansion of the medium, charm flow would lead to a depopulation at very low momentum and further enhance the spectrum at intermediate momentum. Both effects would lead to a nuclear modification factor above unity, with a maximum of $\RAA = 1.2 -1.5$ around $2\gevc$.

Besides parton energy loss, 
initial-state effects might also influence the \RAA measurement. 
In particular, a
modification of the parton distribution functions in the nucleus when compared to the proton alters the initial hard scattering probability, thus
affecting the production yields of hard
partons, including heavy quarks.
In the kinematic range relevant for charm production 
at LHC energies, the main predicted effect is nuclear shadowing, i.e. the reduction of the parton distribution functions for values of $x$ below $10^{-2}$.   
The effect of shadowing on the D meson $\RAA$ was estimated using the 
next-to-leading order framework MNR~\cite{mnr} with the CTEQ6M parton distribution 
functions~\cite{cteq6} for the proton and the EPS09NLO parametrization~\cite{eps09} for the lead nucleus.
The effect from shadowing on the nuclear modification factor 
is estimated to be no larger than $\pm 15\%$ for $\pt>6\gevc$, indicating that 
the strong suppression observed in the data is a final-state effect.

The dependence on color charge and parton mass 
is investigated by comparing
the nuclear modification factor of D mesons and $\pi$ mesons. 
Since final results on the 
pion $\RAA$ at the LHC were not available at the time of publication, a comparison is made
with the measurement of unidentified charged particles which are dominated by pions.
The average D meson nuclear modification factor is close to that of
unidentified charged particles~\cite{chargedRAA}.
However, considering that the systematic uncertainties of D mesons are not fully correlated with $\pt$, there is an indication 
that  $\RAA^{\rm D}>\RAA^{\rm charged}$.
The nuclear modification factor of B mesons 
has been measured by the CMS Collaboration by detecting displaced 
J/$\psi$ mesons with $\pt>6.5\gevc$
stemming from weak decays of B mesons~\cite{CMSquarkonia}.
Their suppression is clearly weaker
than that of unidentified charged particles, while the comparison with D mesons is not yet conclusive.
 
Various models based on calculating parton energy loss may be used to compute the D meson  
nuclear modification factor. Here, a comparison is given between models, which can be divided 
into two groups: 
\begin{itemize}
\item[(A)] models based on microscopic calculations in perturbative QCD,
e.g. (I) - (VIII)~\cite{vitev,vitevjet,whdg2011,horowitzAdSCFT,beraudo,gossiaux,bamps,cujet,adsw}.
Figure~\ref{fig:raa} displays the comparison of these models  
to the average D meson $\RAA$, for central Pb--Pb collisions (0--20\%), along with the 
comparison to the charged-particle $\RAA$~\cite{chargedRAA}, for those models that also compute 
this observable: (I)~\cite{vitev}, (II)~\cite{whdg2011}, (III)~\cite{horowitzAdSCFT}, (VII)~\cite{cujet}. 
While the \RAA values vary drastically between different models, neither model is able to consistently describe the nuclear modification factor for 
light- and heavy-flavor hadrons. The apparent rise in \RAA with transverse momentum is solely due to the shape of the reference \pp spectrum~\cite{Zapp:2012ak}, and not
a particular prediction of any model. 
\item[(B)]
A recent development is the conjectured duality of quantum field theories of e.g. Yang-Mills-Shaw type, describing
elementary particles on one side and theories of quantum gravity, formulated in string-theory, on the other~\cite{maldacena}.
The advantage is that while in quantum field theory the fields are strongly interacting and thus perturbative methods can not be applied, the fields in gravitational theory are weakly interacting, making calculations possible. 
A well studied case is the Anti-de-Sitter/conformal-field theory (AdS/CFT) correspondence. Here, string theory is formulated in an Anti-de-Sitter space with the conformal field theory 
living on the conformal boundary of the Anti-de-Sitter space-time.
One specific example of the AdS/CFT correspondence relates the rather general class of type IIB string theories in ten-dimensional space that is defined by the the product of a five-dimensional Anti-de-Sitter-spacetime 
and a five-dimensional background, $AdS_5 \times S^5$, to  $\mathcal{N}=4$ supersymmetric Yang-Mills theory on the four-dimensional conformal boundary.
The drag force of an external and thus infinitely heavy quark moving in a thermal plasma of a thermalized  $\mathcal{N}=4$ 
supersymmetric Yang-Mills theory has been calculated in such an approach~\cite{Friess:2006fk}.
A model determined drag coefficients (III) by incorporating experimental results from RHIC~\cite{horowitzAdSCFT}  and significantly underestimates the
nuclear modification factor of D mesons. 
\end{itemize}

\begin{figure}[hbtp]
  \centering
  \includegraphics[width=0.6\textwidth]{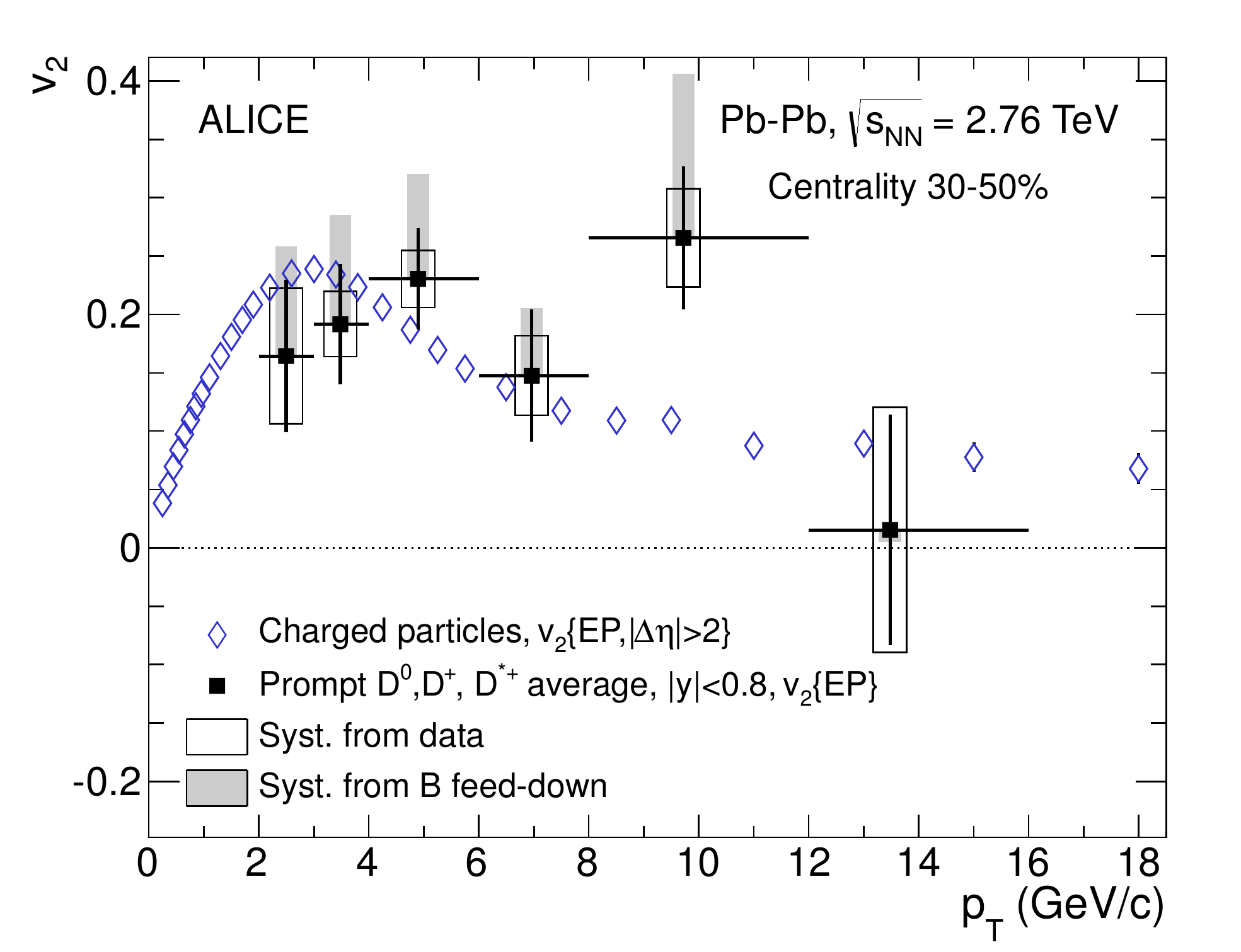}
  \caption{Elliptic flow coefficient $v_2$ for D mesons and unidentified charged particles. This figure has been taken from\cite{dv2}.}
  \label{fig:v2}
\end{figure} 
The measurement of anisotropy in the azimuthal distribution of particle
momenta provides further insight into the properties of the QGP medium.
Anisotropic patterns originate from the initial 
anisotropy in the spatial distribution of the nucleons 
participating in the collision.
The azimuthal anisotropy of produced particles 
is characterized by the Fourier
coefficients $v_n=\langle\cos[n(\varphi-\Psi_n)]\rangle$,
where $n$ is the order of the harmonic, $\varphi$ is the azimuthal angle of 
the particle, and $\Psi_n$ is the azimuthal angle of the initial state 
symmetry plane for the $n$-th harmonic.
For non-central collisions the overlap region of the colliding nuclei 
has a lenticular shape 
and the azimuthal anisotropy 
is dominated by the second Fourier coefficient $v_2$, commonly denoted by 
elliptic flow~\cite{ALICEv2, ALICEhighharm}.

Two mechanisms are responsible for generating a non-zero $v_2$.
The first mechanism, dominant in the bulk, e.g. at low ($\pt<3\gev/c$) and intermediate 
(3--6$\gevc$) transverse momentum, is the build-up of a collective expansion 
through interactions among the medium constituents.
An anisotropic component in this collective expansion develops mainly in
the early stages of the lifetime of the system, when the spatial anisotropy is 
large~\cite{Ollitrault,hydro,Heinz}.
The second mechanism 
is the path-length dependence of in-medium parton energy loss, due to
medium-induced gluon radiation and elastic collisions. 
This differential nuclear modification with respect to the reaction plane is predicted to give rise to a positive $v_2$ for 
hadrons up to large transverse momenta~\cite{highptv2a,highptv2b}.
The $v_2$ values measured for light-flavor hadrons at RHIC and LHC 
energies can be described for the low-$\pt$ region in 
terms of collective expansion of a strongly-interacting 
fluid~\cite{STARv2,hydroVisc,ALICEv2,CMSv2}, and for the high-$\pt$ region 
($\pt>$6--8$\gev/c$) in terms of path-length dependent parton energy
loss~\cite{ATLASv2,CMShighptv2,ALICEhighptv2,whdg2011}.

The measurement of the elliptic flow of charmed hadrons provides further
insight into the transport properties of the medium.
At low $\pt$, charmed hadron $v_2$ offers a unique opportunity to test whether 
quarks with large mass ($m_{\rm c}\approx 1.5\gevcc$) also
par\-ti\-ci\-pa\-te in the collective expansion dynamics and possibly 
thermalize in the medium~\cite{Batsouli,GrecoKoRapp}.
Because of their large mass, charm quarks are expected to have a longer
relaxation time, i.e.\ time scale for approaching equilibrium with the 
medium, with respect to light quarks~\cite{Teaney}.
At low and intermediate $\pt$, the D meson elliptic flow is expected to
be sensitive to the heavy-quark hadronization mechanism.
If there are substantial interactions with the medium, a significant 
fraction of low- and intermediate-momentum heavy quarks could hadronize via 
combination with other quarks from the bulk of thermalized 
partons~\cite{Molnar,GrecoKoLevai}, thus enhancing the $v_2$ of D mesons 
with respect to that of charm quarks~\cite{GrecoKoRapp}.
In this context, the measurement of D meson $v_2$ is also relevant for the 
interpretation of the results on $\jpsi$ anisotropy~\cite{jpsiv2}, since 
$\jpsi$ mesons formed from ${\rm c\overline{c}}$ combination would inherit 
the azimuthal anisotropy of their constituent 
quarks~\cite{LiuXuZhuang,ZhaoEmerickRapp}. 
At high $\pt$, the D meson $v_2$ can constrain the path-length dependence of
parton energy loss, complementing the measurement of the suppression of 
particle yields with respect to the expectation from proton--proton 
collisions. 

Theoretical models of heavy-quark interactions with the medium constituents 
enable the calculation of both the $v_2$ and $\RAA$ of heavy-flavor mesons in a 
wide $\pt$ range~\cite{gossiaux2,bamps2,rapp2,beraudo,Lang}.
For semi-central collisions at the LHC, they predict a large elliptic 
flow ($v_2\approx 0.1$--0.2) for $\rm D$ mesons at $\pt\approx 2$--3\gevc 
and a decrease to a constant value $v_2\approx 0.05$ at high $\pt$.

The azimuthal anisotropy in heavy-flavor production was measured 
in Au--Au collisions at $\rtsnn=200\gev$ at RHIC using electrons from 
heavy-flavor decays.
The resulting $v_2$ values are as large as 
0.13~\cite{phenixHFEv2,phenixHFERAAv2}.

For D mesons, the second Fourier coefficient, $v_2$, was calculated
 according to:
\begin{equation} \label{eq:v2}
v_2=\frac{1}{R_2}\frac{\pi}{4}\frac{N_{\rm in\mbox{-}plane}-N_{\rm out\mbox{-}of\mbox{-}plane}}{N_{\rm in\mbox{-}plane}+N_{\rm out\mbox{-}of\mbox{-}plane}}\,.
\end{equation}
Due to the limited statistics, the integrated asymmetry between the yields in the azimuthal range from $\phi = 0-90^\circ$ and $\phi = 90^\circ-180^\circ$ was considered.
The factor $\pi/4$ results from the integration of the
second term, $2\,v_2\cos(2\Delta\varphi)$, of the $\dNdphi$ distribution
in the considered intervals of relative azimuth.
The factor $1/R_2$ is the 
correction due to the angular resolution in determining the symmetry plane 
$\Psi_2$~\cite{PoskVol}, which smears out the angular modulation, leading to an apparent lower value for the observed $v_2$.

The resulting D-meson $v_2$ is shown in Fig.~\ref{fig:v2}.
The average of the measured $v_2$ values in the interval $2<\pt<6\gev/c$ is
$0.204\pm 0.030$~(stat)~$\pm 0.020$~(syst)~$_{-0}^{+0.092}$~(B feed-down),
which is larger than zero with $5.7\,\sigma$ significance.
A positive $v_2$ is also observed for $\pt>6\gev/c$, which most likely 
originates from the path-length dependence of the partonic energy loss, 
although the large uncertainties do not allow a firm conclusion.
The measured D meson $v_2$ is comparable in magnitude to that of charged 
particles, which are dominated by light-flavor hadrons~\cite{ALICEhighptv2}.
This consistency suggests that the relaxation time of charm quarks in the 
medium is similar to that of light partons and it is short with respect to the 
time scale for diluting the initial geometrical anisotropy, possibly
indicating that low momentum charm quarks take part in the 
collective motion of the system.
The measured $v_2$ tends to favour the models that predict a larger 
anisotropy at low $\pt$~\cite{gossiaux2,bamps2,rapp2,Lang}.
Despite the relatively large uncertainties, these measurements on the nuclear modification factor and elliptic flow 
for D mesons together present the most stringent constraints  
to date on models describing the interaction of charm quarks with the QCD medium.

\newpage
\subsection{Correlations of charm and anti-charm quarks}
\label{section:model}

The charm and beauty quantum numbers are conserved in strong interactions. Therefore,
in strong interactions, heavy quarks are always created together with their anti-quark and are thus correlated.
In collisions of leptons it has been shown that a hadron containing a charm or beauty quark carries a 
significant fraction of the initial quark momentum~\cite{aleph_95,delphi_95,opal_95}.
Hence, initial heavy-quark correlations survive the fragmentation
process into hadrons to a large extent, and are observable e.g. in the angular distributions 
of pairs of $D$ and $\overline{D}$ mesons~\cite{Lourenco,e791}. 
In pp collisions, 
the experimentally observed correlations of D mesons, measured at
fixed target experiments~\cite{Lourenco} are reproduced well 
by the Monte Carlo event generator 
PYTHIA~\cite{Sjostrand:2000wi}.

\begin{figure}[hbtp]
  \centering
  \includegraphics[width=0.54\textwidth]{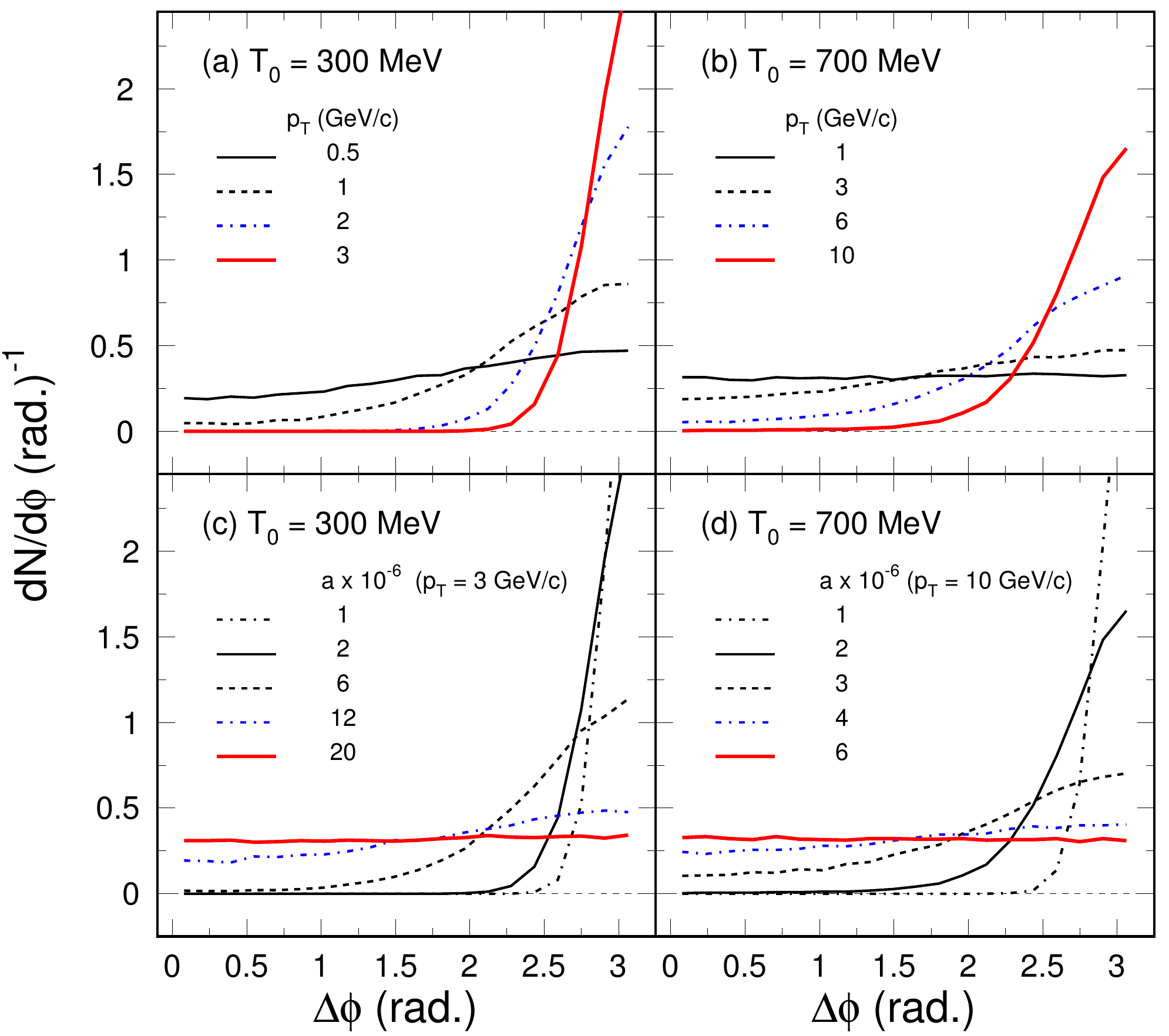}
  \caption{Correlations in relative azimuth $\Delta \phi$
  of \DDbar\ pairs from Langevin calculations with
  $T_0=300$\mev (RHIC) and $700$\mev (LHC).  The upper part shows the
  dependence of the correlations on the initial $p_{\rm T}$ of the charm quark; 
  the lower part shows the drag
  coefficient dependence. This figure has been taken from~\cite{Zhu}.}
  \label{fig:corr}
\end{figure}

In high-energy collisions of heavy nuclei, frequent interactions among partons (quarks and gluons) of the 
medium and heavy quarks
may lead to a significant modification of these initially existing correlations.
On the other hand, hadronic interactions at the late stage are insufficient
to alter the azimuthal correlations of 
$D\overline{D}$ pairs~\cite{Zhu}. 
Frequent interactions
distribute and randomize the available (kinetic) energy and finally drive the system, 
i.e. light quarks and gluons,
to local thermal equilibrium.
To what extent this also happens for heavy quarks is currently a subject of discussion~\cite{Moore:2004tg,rapp,Cao:2011et,He:2011yi,Yan:2007zze}.
A decrease in the strength of heavy quark correlations in high-energy collisions of heavy nuclei 
as compared to \pp collisions
would indicate early thermalization also of heavy quarks.

We studied the modification of azimuthal correlations of 
D and $\overline{\rm D}$ meson pairs as a sensitive indicator of
frequent occurrences of partonic scattering. 
To explore how the QCD medium generated in central ultra-relativistic
nucleus-nucleus collisions influences the correlations of D and
$\overline{\rm D}$ meson pairs, we employed a non-relativistic Langevin approach
which describes the random walk of charm quarks in a QGP and was first
described in
Refs.~\cite{Svetitsky:1996nj,Svetitsky:1997xp,Svetitsky:1997bt},
\begin{equation}
\frac{d\vec{p}}{dt} =-\gamma(T)\vec{p}+\vec{\eta}(T),
\end{equation}
with the drag coefficient $\gamma$ and the normalized Gaussian noise variable  $\vec{\eta}$ describing the heavy quark diffusion (random walk).
The drag coefficient $\gamma$ quantifies the resistance of the heavy quark in the QGP fluid. A large drag coefficient would lead to large heavy-quark flow.
Both parameters are dependent on the local temperature of the system and were parameterized as in Ref.~\cite{Svetitsky:1996nj},
e.g $\gamma = a T^2$,
neglecting any momentum dependence of the drag coefficient $\gamma$. They are related through the fluctuation-dissipation relation in equilibrium
assuming a charm quark mass of $1.5\gevcc$.

In order to isolate the effects purely caused by parton-parton rescattering in the
medium, charm and  anti-charm quarks were generated with the same $p_{\rm T}$ and zero longitudinal momentum back-to-back, i.e.\ at
relative azimuth $\Delta \phi = \pi$,
and  a delta function was used to fragment the
charm quark into a charmed hadron at the hadronization stage. 

Results for D meson (charm
quark) pairs on their angular correlations 
are shown in Fig.~\ref{fig:corr}\,(a)
for different initial charm quark $p_{\rm
T}$ values, and  $T_0=300$\mev, values typical for RHIC collisions. 
The fastest charm quarks, represented by the highest momenta, are able to escape from the QGP without suffering
significant medium effects, while the slower quarks, represented by the lowest momenta, have their pair azimuthal correlation
almost isotropic due to interactions in the medium.

Results for a higher initial temperature $T_0=700$\mev, representative of LHC energies, 
are shown in 
Figure~\ref{fig:corr}\,(b).
Here, interactions of
charm quarks with the medium are so frequent that only the most
energetic charm quarks preserve part of their initial angular
correlation; low $p_{\rm T}$ pairs are even completely kinetically equilibrated
and become part of the medium. 
Note that  in the actual calculations, only longitudinal flow is considered. 
In reality, transverse radial flow
leads to an enhancement of the same-side correlation~\cite{Voloshin:2003ud} and will be present especially
when interactions are frequent~\cite{Moore:2004tg,Nahrgang:2013saa,Adare:2013wca}.

The above results were obtained with a drag coefficient estimated with
perturbative QCD adopting a large coupling constant $\alpha_s=0.6$~\cite{Svetitsky:1987gq}. 
Recent pQCD calculations~\cite{GolamMustafa:1997id,hees_05} show
a factor of 2-3 smaller drag coefficient. However, non-perturbative contributions that arise from
quasi-hadronic bound states in the QGP might be important~\cite{hees_05}.
This would result
in a much larger drag coefficient. Since exact values of the drag
coefficient from lattice QCD calculations do not exist, the
sensitivity on variations of the parameter $a$ within
the given range is considered.  Figures~\ref{fig:corr}\,(c) and (d) show the drag
coefficient dependence of the charm anti-charm angular correlations
for high-energy charm quarks: $p_{\rm T}=3(10)\gevc$ and $T_0=300(700)\mev$. 
The correlations
dissipate when $a$ is increased by around a factor of 5(2)
with respect to the pQCD value.

\begin{figure}[hbtp]
  \centering
  \includegraphics[width=0.54\textwidth]{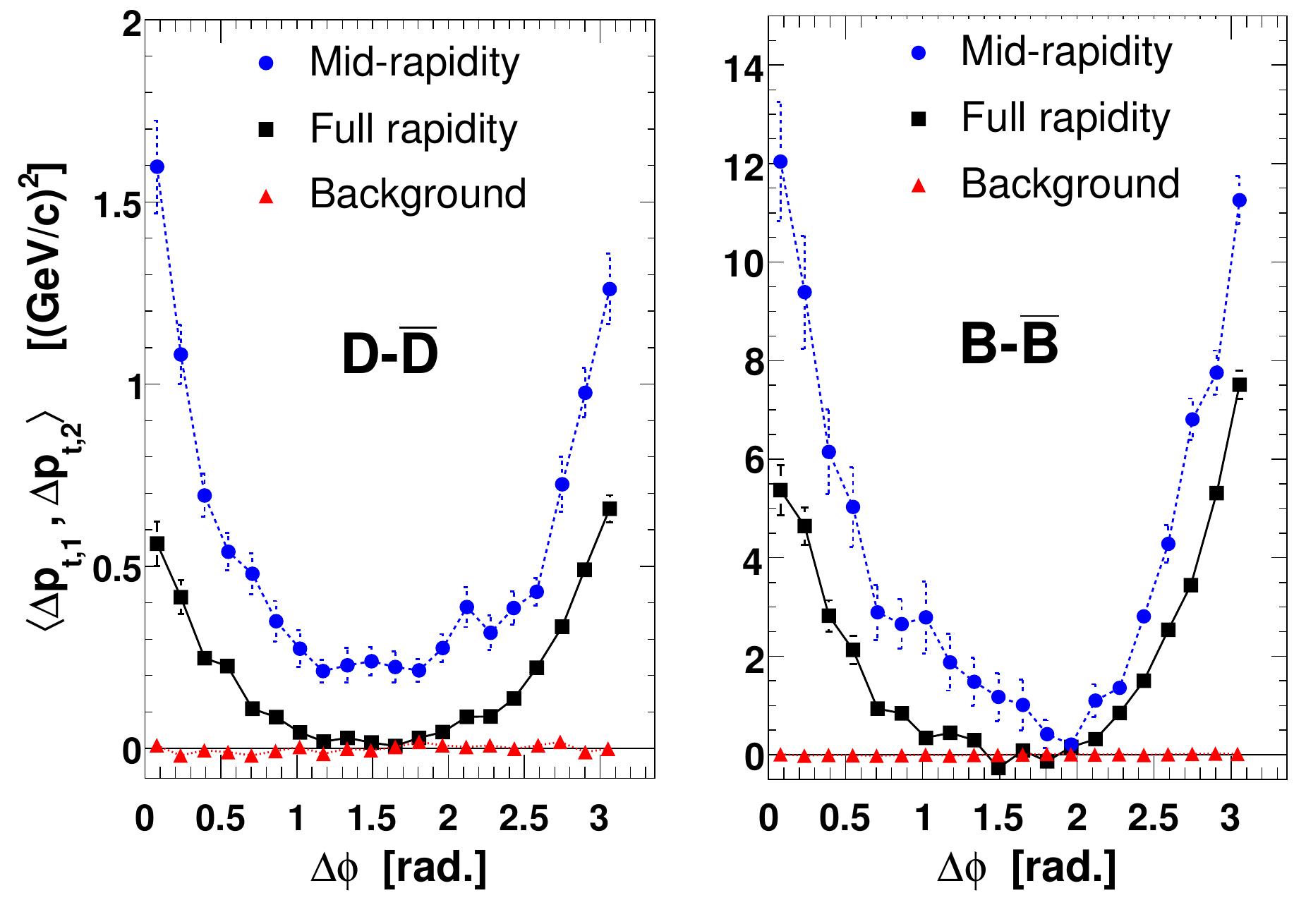}
  \caption{(Color online) Distribution of  the momentum correlator $\corr$ of $5 \cdot 10^5$ $D\overline{D}$ pairs (left panel) and 
$5 \cdot 10^5$ $B\overline{B}$ pairs (right panel) 
as a function of relative azimuth $\Delta\phi$ 
at mid-rapidity~(circles), for full rapidity~(squares) and for background using the mixed event method (triangles) 
for $pp$ collisions at $\sqrt{s}$ = 14\tev as calculated using PYTHIA (6.406). This figure has been taken from~\cite{Tsiledakis:2009qh}.}
  \label{fig:cor2}
\end{figure}


At LHC energies, higher-order processes become dominant, and
sensitivity to heavy quark thermalization might be lost. 
To overcome this complication, a two-particle transverse momentum correlator 
$\corr$ was applied to pairs of  heavy-quark hadrons
and their semi-leptonic decay products as a function of their relative azimuth.
This has the following advantages: \newline
(i) The correlator is sensitive to non-statistical fluctuations, thus scraping out any physical correlation.\newline
(ii) In case of physically uncorrelated candidate-pairs (e.g. combinatorial background), the extracted value for the correlator vanishes,
thus providing a reliable baseline.\newline
(iii) A localization of the observed correlations in transverse momentum space may help to obtain further insight into
the origin of the observed correlations in relative azimuth.

Calculations at leading order (LO)  contain flavor
creation processes ($q\overline{q}\rightarrow Q\overline{Q}$, $gg\rightarrow Q\overline{Q}$)
and lead to an enhancement at relative azimuth around $\Delta\phi \approx 180^{\rm o}$ , i.e. 
the $\rm{D}$ meson pair is preferentially emitted back-to-back. 
In flavor excitation processes
($qQ\rightarrow qQ, gQ\rightarrow gQ$), one heavy quark from the proton sea participates in the hard scattering,
leading to an asymmetry in momentum and the relative angular distribution.
In gluon splitting processes ($g\rightarrow Q\overline{Q}$), one gluon splits into a heavy-quark pair with relatively small opening angle and
transverse momentum. At even higher orders, the distinction becomes scale dependent and is thus less well-defined.

In order to extend calculations in PYTHIA beyond leading order, processes contributing at higher orders
were calculated using a massless matrix element~\cite{ALICE_PPRII} applying 
a lower cut-off in the transverse momentum-transfer scale of the underlying hard scattering to avoid divergences in
the calculated cross section~\cite{ALICE_PPRII}.   
The PYTHIA parameters were subsequently tuned to reproduce these next-to-leading order predictions~\cite{ALICE_PPRII}.
For the hadronization of charm and beauty quarks 
the Lund fragmentation scheme was used~\cite{pythia2}.  

The correlator 
for pairs of \DDbar\ mesons (left) and \BBbar\ mesons (right) as shown in Fig.~\ref{fig:cor2}
has a pronounced forward-backward peaked structure.
We observe an enhancement at small azimuth from gluon
splitting processes, while flavor creation of $c\overline{c}$-quark pairs leads to an enhanced  correlation at backward angles.
We have checked that flavor excitation processes lead to a rather flat distribution. 
To mimic combinatorial background, which is always present in the experiment, we applied the correlator 
to  $D$ and $\overline{D}$ mesons  from different \pp collisions, which are physically uncorrelated. 
This results in a value  consistent with zero (see Fig.~\ref{fig:cor2}). Therefore the correlator
allows for a clear distinction between the case where correlations are present (different from zero) or absent (equal to zero)

However, the full kinematic reconstruction of $D$ mesons from topological decays suffers from small branching ratios and 
rather small reconstruction efficiencies, resulting in low statistics.
This is especially the case when pairs of  $\rm D$ mesons are considered, as in this case where these factors enter quadratically.
As an alternative, we considered electrons (positrons) from the semi-leptonic decays of
charm and beauty hadrons,
with an average
branching ratio to electrons of 10\% and 11\%, respectively.
This clearly indicates that the initial correlations among a 
heavy quark and its corresponding anti--quark even survive 
semi-leptonic decays into electrons (positrons) to a large extent.

Further, a comparison of experimental heavy-quark correlations from \pbpb\ collisions to results from microscopic transport calculations 
would provide an independent way to extract effective heavy-quark scattering rates in the QCD medium~\cite{molnar07,He:2013zua}. 
Thus, it might be possible to extract general transport properties, which in turn provide important insight into the microscopic in-medium properties
of partons in the QGP and thus the nature of the plasma itself.  However, this information would be lost if heavy quarks fully equilibrate with 
the light partons in the medium. 

On the other hand, open heavy-quark correlations also become important in the study of dilepton invariant-mass spectra
in high-energy nuclear collisions since the contribution of correlated open heavy-quark decays competes with dilepton emission rates
from the QGP in the intermediate invariant-mass range~\cite{ceres02, PHENIX2010,Xu:2013uza,Lang:2013wya}. Thus, solid experimental constraints on the extent of the loss of
open heavy-quark correlations are essential for an interpretation of the dilepton invariant-mass spectra, in particular concerning mechanisms 
of chiral symmetry restoration in the hot partonic medium created in high-energy nuclear collisions at LHC energies.

Recently, the CMS Collaboration has measured angular correlations of \BBbar\ mesons with momenta above 15\gevc in \pp collisions at \rts=7\tev and found
a dominant contribution from gluon splitting processes which is underestimated by calculations in perturbative QCD~\cite{Grab:2011ta}.

\newpage
\section{Outlook}
\label{section:outlook}
Heavy flavor hadrons carrying a charm or beauty quark are unique probes in modern nuclear and particle physics. Their large mass allows for a quantitative description of their production within perturbative calculations in QCD. A good agreement is found between experiment and theory, within rather large uncertainties in both. Hadronization of charm allows for a description within the Statistical Model. Present experimental results might hint 
that the contribution from gluon splitting processes to heavy quark production is too low. This might be cured by using a smaller charm mass than 
is currently assumed in the calculations. 

Lead-lead collisions at the highest energies show that charm quark interactions with the QCD medium are substantial.   
Precise measurements of the phase space distribution of charmed hadrons will address the open question of whether the predicted mass hierarchy of quark energy
loss in the hot and dense QCD medium is realized in nature. This addresses the issue of in-medium parton dynamics,  which is not yet understood at the
fundamental level. Closely related is the medium response to the energy deposit from a heavy quark. The ultrahigh-resolution vertex detectors at the LHC  make identification of heavy-quark jets on an event-by-event basis possible. This gives unique access to quark jets, in contrast to gluon jets, which dominate high-momentum particle production at the LHC.  At the low momentum side, precise measurements of heavy-quark flow and correlations will
decisively address  the question of the extent to which heavy quarks participate in the collective expansion of the bulk or even reach kinetic equilibrium. 
It will then be possible to extract general transport properties, which give an important insight into the microscopic in-medium properties of partons in the QGP, and thus the nature of the plasma itself. 
However, this information would be lost in the case of heavy quarks fully equilibrating with the light partons in the medium.

These considerations make detection of heavy-quark hadrons down to zero momentum essential, as the topological selection of heavy-quark decays becomes
far more challenging in this region. Improved secondary vertexing, in conjunction with excellent particle identification 
on a track-by-track basis and sophisticated approaches combining information on particle identification from different detectors, e.g. using a Bayesian approach, will help to reduce the combinatorial background and thus extend the accessible momentum range to the lowest momenta. This will also help in detecting charmed baryons, e.g. $\Lambda_c^+$, which have much shorter lifetimes than the charmed mesons. Also, the relative abundances of charmed hadrons might significantly differ in lead-lead collisions compared to a charm hadronizing in vacuo. In particular, the lifting of strangeness suppression in central lead-lead collisions might alter the role of the strange charmed meson $D_s^+$~\cite{rapp2}. 
Full reconstruction of low-momentum B mesons in central lead-lead collisions is limited by low production yields and per mille branching ratios. This might be overcome by yet-increased luminosities being achieved by the accelerator, in combination with faster readout-rate capabilities in the experiments at the LHC.

These goals are within reach in two steps:
After the first long shutdown of the LHC, a ten-fold increase in statistics is expected for the second data-taking period starting in 2015. 
During the second long shutdown foreseen for the whole year of  2018, all experiments will undergo major detector upgrades, including vastly
improved secondary-vertexing and readout-rate capabilities. Another increase in statistics of a factor of ten (100 in total) is expected at the
high-luminosity LHC. 

Complementary to this, charm production close to the threshold and its implications on superdense baryon-rich QCD matter will be studied  in (anti)proton-proton, (anti)proton-nucleus and nucleus-nucleus collisions at the Facility for Antiproton and Ion Research (FAIR) currently under construction at the GSI Helmholtz Centre for Heavy-Ion Research, Darmstadt, Germany~\cite{cbm,Lutz:2009ff}.

\newpage
\bibliographystyle{href-physrev4}   %
\bibliography{habil}
\newpage
\end{document}